# A Dynamic Simulation-Optimization Model for Adaptive Management of Urban Water Distribution System Contamination Threats

Amin Rasekh; Kelly Brumbelow

Zachry Department of Civil Engineering, Texas A&M University

Corresponding author:   Amin Rasekh – Postdoctoral Research Associate

Zachry Dept. of Civil Engineering, Texas A&M University

3136 TAMU, College Station, TX 77843-3136, USA

**Abstract**. Urban water distribution systems hold a critical and strategic position in preserving public health and industrial growth. Despite the ubiquity of these urban systems, aging infrastructure, and increased risk of terrorism, decision support models for a timely and adaptive contamination emergency response still remain at an undeveloped stage. Emergency response is characterized as a progressive, interactive, and adaptive process that involves parallel activities of processing streaming information and executing response actions. This study develops a dynamic decision support model that adaptively simulates the time-varying emergency environment and tracks changing best health protection response measures at every stage of an emergency in real-time. Feedback mechanisms between the contaminated network, emergency managers, and consumers are incorporated in a dynamic simulation model to capture time-varying characteristics of an emergency environment. An evolutionary-computation-based dynamic optimization model is developed to adaptively identify time-dependant optimal health protection measures during an emergency. This dynamic simulation-optimization model treats perceived contaminant source attributes as time-varying parameters to account for perceived contamination source updates as more data stream in over time. Performance of the developed dynamic decision support model is analyzed and demonstrated using a mid-size virtual city that resembles the dynamics and complexity of real-world urban systems. This adaptive emergency response optimization model is intended to be a major component of an all-inclusive cyberinfrastructure for efficient contamination threat management, which is currently under development.

# 1. Introduction

Safe and reliable drinking water is vital to every community. Approximately 90% of the U.S. population receives water from one of 170,000 public water utilities. These drinking water distribution systems (WDSs) are inherently vulnerable to natural or malicious contamination because of their ubiquity, multiple points of access, the aging infrastructure, and the increased risk of terrorism. The plethora of past contamination incidents and foiled attempts further demonstrates the vulnerability of municipal water supply systems [1, 2]. Title IV of the Public Health Security and Bioterrorism Preparedness and Response Act of 2002 [3] mandates all community water systems that serve more than 3,300 people to prepare or revise emergency management plans.

WDS contamination emergency management is based on risk assessments and includes four major phases of hazard mitigation, emergency preparedness, emergency response, and disaster recovery [4]. This study is particularly more focused on emergency response phase. A contamination emergency response phase is initiated when an actual (or potential) injection of contaminant is propagated (or will propagate) over a WDS, and it extends until the situation is stabilized (e.g., when the risk of health consequences has returned to normal conditions). Emergency response decision aid tools have been devised in previous studies to assist water utility operators in making better decisions for public health protection against contamination events. Genetic algorithms (GA), decision trees, and heuristic approaches have been used for guiding contaminant containment and flushing operations [5-7]. Agent-based models have been developed and coupled with a GA to optimize routing of siren vehicles to best warn and protect consumers from exposure [8, 9]. Multiobjective simulation-optimization has been applied to help utility operators achieve conflicting response objectives such as health protection and WDS service interruption prevention [10-14].

Previous studies have generally applied static optimization approaches to find optimal response protocols on the implicit assumption that the optimization problem is fixed during the course of an emergency. From an optimization perspective, this assumption implies that the response optimization fitness functions (e.g., minimization of ultimate health impacts) are temporally constant. In other words, fitness of a particular response protocol is not changing during the optimization process once it is started.

In reality, however, the fitness functions are feedback-influenced by many uncertain and dynamic factors, i.e., projected spread of contaminant and actions taken by the utility operators and consumers [15, 16]. Under the effect of these unpredictable factors, the search domain may repeatedly change over time, and the current best response protocol might be no longer optimal moments later. The decision support model should thus explicitly account for this unpredictably time-varying system behavior to accurately reproduce the reality and identify best contamination impact reduction decisions in a timely manner.

Dynamic optimization has been successfully applied in different disciplines for solving optimization problems in changing environments. Dynamic optimization methods methodically exploit and transfer useful knowledge from older environments and maintain adaptability to guide and speed up the exploration in emergent environments [17, 18]. Some recent successful applications include products pricing [19], contaminant source characterization [20], vehicle routing [21], military mission planning [22], electric power supply optimization [23], video-based face recognition [24], and dynamic traveling salesman problem [25]. Despite this rich record of successful applications, however, dynamic optimization has not been employed to solve the crucial problem of dynamic water contamination emergency management yet. This study focuses on this unexplored research area and develops dynamic simulation-optimization models for adaptive management of drinking water networks contamination emergencies.

We investigate the application of dynamic modeling for adaptive emergency response to disasters in the scope of drinking WDS contamination events. A dynamic simulation-optimization scheme is developed to identify and track time-dependant optimal response protocols to provide emergency managers with adaptive decision support in real-time. The adaptive simulation model accounts for multiple uncertainty factors that contribute to the unpredictable time-varying system behavior, including time-dependant perceived contamination source attributes, consumers' actions, and emergency response operations. This dynamic optimization scheme uses an evolutionary-computation-based multiobjective approach, which methodologically preserves diversity in the search process for enhanced adaptation against the effect of changes.

In what follows, the structure and elements of the dynamic simulation model developed to simulate the dynamic emergency environment are first described. This is followed by the formulation of the time-varying optimization fitness function and decision variables and a brief overview of existing dynamic optimization techniques. The evolutionary-computation-based algorithm used here for the dynamic optimization of health protection response protocols is then described. The proposed adaptive simulation-optimization model is then evaluated and discussed on a virtual city, Mesopolis, which possesses the spatial and temporal complexity of real-world cities. The results are analyzed to illustrate the system behavior during an emergency and investigate model performance. The paper concludes with a number of recommendations for enhancing applicability and efficiency of the proposed decision support model in the future.

## 2. Dynamic Emergency Environment Simulation

A WDS contamination emergency threat starts with an actual (or potential) release of contaminant that propagates across the network, and it extends until the risk of health impacts has returned to pre-event levels. Under an unfolding contamination threat, the system might exhibit an uncertain and irregular behavior that could radically deviate from normal operation conditions. Knowledge of contaminant source attributes that dictate emergency response decisions evolves as more information streams in over time. As the event progresses, emergency managers take response actions based upon their current knowledge of the state of the system and these actions change the state of the contamination in the WDS and alert consumers. Alerted and exposed consumers may subsequently alter their water consumption choices, which consequently affect network hydraulics and contaminant plume spread [15]. Conceptualizing and modeling these different sources of variability is crucial for realistic simulation of system behavior and effective reduction of contamination impacts.

This section first defines a contamination scenario and how the emergency managers' perception of an occurred scenario evolves as more information streams in over time. Then, it describes the two particular emergency response mechanisms considered in this study (i.e., contaminant flushing through opening hydrants and public warning through food-grade dye injection) and how their implementation contributes to the system dynamics. It is followed by a description of consumers' behavior during a contamination emergency and its influence on the state of the contamination and health consequences. The last part of this section describes the hydraulic model used in this work to simulate WDS hydraulics and contaminant propagation during a contamination event.

2.1. Perceived Contamination Scenario Updates

A contamination scenario is defined by a set of attributes (e.g., contaminant injection location and duration) and corresponds to a specific level of ultimate consequences [26]. The attributes of a contamination scenario are estimated through integrated assessment of different system observations and evidence. This information may stream from physical security alarms, supervisory control and data acquisition (SCADA) system, and consumers' complaints. Bayesian and optimization models can be applied to process the streaming data in real-time and update estimated scenario attributes using existing and incoming information [20, 27-30].

Since the perceived contamination scenario dictates the effectiveness of the health protection strategies taken, the optimization process needs to adapt to scenario updates to be capable to continuously track the optimum in a time-varying search space. Perceived scenario attributes are thus treated here as time-varying parameters opposed to past studies that assume the perceived contamination scenario is fixed while the optimization is being performed. It should be emphasized that the *true* contamination scenario has already occurred and is thus not time-varying, but it remains unknown to the emergency managers. During the course of an emergency, the streaming data is processed adaptively at every stage and perceived contamination scenario is updated to better estimate the true scenario.

Development of an adaptive scenario characterization model is not the focus of this study and the proposed dynamic response optimization model is thus demonstrated on an arbitrary scenario update timeline without loss of generality. The adaptive simulation-optimization scheme proposed here should be ideally coupled with a dynamic source identification model [20] that continuously updates the perceived contamination scenario using streaming data. A general framework for dynamic integration of these analytical modules for development of an adaptive cyberinfrastructure for threat management is described in [31].

2.2. Water Utility Operations

In the event that a contaminant is introduced to a WDS, water utility operators may take different preventive and protective actions to protect public health. Preventive actions operate on the system to reduce impacts and may include discharging contaminated water through opening fire hydrant or isolating contaminated areas via closing valves. Protective actions require action by the consumers to protect themselves from contaminant exposure and might consist of broadcasting various protective action recommendations such as boil-water advisory through the media [32].

The decisions on timing and frequency of response actions are made with consideration of threat observations credibility and unintended adverse response side effects (e.g., system service interruptions and false public warnings). The actions taken by the utility managers would change the normal hydraulic conditions, and thus the propagation of the contaminant plume in the system. Implementation of response actions thus alters ultimate public health consequences and affects efficacy of future response decisions. Moreover, effectiveness of consequence reduction measures expectedly degrades as time passes due to wider contaminant spread and prolonged public exposure to the contaminant. The simulation model developed here adaptively reevaluates response actions considering the increasing response execution time delay and the effect of actions that are already executed or being executed on system behavior. Without loss of generality, two particular response mechanisms of hydrant operation for contaminant flushing and food-grade dye injection for public warning are included in the model.

Contaminant flushing is most commonly performed through opening fire hydrants to discharge a large volume of contaminated water from the system. Flushing operations should be arguably performed at locations where contaminant concentration is relatively higher. Otherwise, it may possibly worsen the consequences through unintentionally diverting the contaminant plume to uncontaminated regions [United States Environmental Protection Agency [33, 34]. This urges utilization of a methodological optimization model to inform contaminant flushing decisions and prevent inadvertent increase in health consequences.

Food-grade dye injection is a novel emergency response mechanism for WDS contamination consequence management [33]. It acts as a warning mechanism that discourages public consumption of contaminated drinking water and has been computationally demonstrated as a potentially effective response mechanism for public health protection [14]. Allura Red dye (also known as "Red 40" and "E129") is chosen in this study among different available food-grade dyes because it has an intense red color and is approved by the U.S. Food and Drug Administration [35]. Food-grade dye injection has certain inadvertent side consequences such as false public alarms and ruined laundry and thus should be implemented when contamination trigger events are credible enough. A complete description of this response strategy and a detailed discussion on its advantages and limitations are provided in Rasekh et al. [14] and is thus not repeated here.

2.3. Consumer Behavior

Different consumers ingest varying amounts of contaminant depending upon time-dependant concentration of the contaminant and dye in their tap water, and the water consumption choices they make under an unfolding contamination incident. The exposure model used here assumes consumers ingest tap water at the typical starting times for the three main meals (7:00, 12:00, and 18:00) and times halfway between these meals (9:30 and 15:00). Alternative probabilistic ingestion timing models [15, 36] may be used to more accurately reproduce the real ingestion behaviors. The tap water ingestion rate for every consumer is set to 0.93 L/day here based upon comprehensive studies performed by USEPA [37].

Consumers cease drinking tap water when they become aware that their tap water is contaminated. This happens when they are sickened and assume tap water is the cause or when they observe intense dye color in their tap water. A series of rules needs to be defined for modeling consumers' reactions and their water usage changes. A consumer will experience symptoms once a threshold dose (i.e, toxic dose (TD)) has been ingested. It is assumed that the exposed consumer changes water usage patterns and stop drinking water within a time period, which is set to one hour here, after TD has been consumed. The contaminant agent used here is arsenic with minimum, average, and maximum estimated TD of 2.45, 3.5, and 4.97 mg for a body weight of 70 kg [38]. Consumers may also become alerted through observation of dye. It is presumed that they become alerted and cease drinking water for the rest of contamination incident once the dye concentration exceeds a relatively high threshold. This threshold is set here to the concentration of Allura Red in commercial soft drinks – roughly 25 mg/L as reported by Lopez-de-Alba et al. [39]. The simulation model checks this observation of high-intensity dye only at the daily ingestion times described in the time-of-ingestion model.

These water usage reduction choices made by the alerted consumers influence the hydraulic state of the network, and thus the spread of the contaminant plume in the system. Consumers may suspend contact uses, such as hand washing, dishwashing, and bathing after they become alert to the contamination. Other non-consumptive uses, however, may continue, such as toilet flushing, landscape watering, and pipe leaks. Such uses are assumed to comprise on average 60%, 51%, and 43% of the total demand for low, medium, and high density residential demands, respectively, using the information reported by Vickers [40] for urban water use. It is assumed the residential users reduce their water usage to these fractions after they are alerted. Industrial users are assumed to maintain 96% of their total water usage. A more complex model may be developed to more realistically reproduce system's behavior through incorporation of consumers' mobility and word-of-mouth communication [15, 41-43]. A more realistic modeling, however, would be often achieved at the price of longer simulation time.

2.4. WDS Hydraulic Simulation

EPANET software is used for hydraulic simulation of the WDS. It is an open-source hydraulic and water quality modeling program developed by USEPA [44]. EPANET provides an integrated computational environment for an extended-period hydraulic and quality simulation of WDSs within pressurized pipe networks.

Demand-driven analysis (DDA) hydraulic modeling approach is used in this study. A basic premise of DDA models is that nodal water demands are fully satisfied during the simulation so that nodal head and pipe flow rates can be estimated through solving a system of quasi-linear equations [45]. This premise holds true in this study due to the fact that considered response strategies of hydrant operation and food-grade dye injection do not cause significant increase in demands or disconnection of demand nodes from pressurized water supply. No pressure-deficit conditions were also occurred during model application phase. Implementation of other response strategies (e.g., valve closure), however, may lead into abnormal pressure-deficit conditions where accuracy of DDA results is questionable. Under such conditions, replacement of the DDA network solver with a pressure-driven solver [13, 46] is recommended.

EPANET toolkit may be coupled with EPANET multi-species extension [47] to more realistically simulate contaminant and dye propagation in the network. This is performed through consideration of decay, density effects, and reaction with wall materials and other dissolved species. In this study, these detailed aspects are not considered and the contaminant and dye transport in the network are modeled as a perfect tracer to avoid associated extra computational time. Besides, there might be noticeable dissimilarities between different preset parameters (e.g., nodal demands) of hydraulic model and the real system during the course of an event. Real-time data provided by the SCADA system can be systematically communicated with the hydraulic model to calibrate and update its parameters [48].

## 3. Dynamic Evolutionary Optimization

The optimum of a time-dependant optimization problem is a time-varying solution. Therefore, solving a time-dependant optimization problem requires optimization algorithms to not only search for the optimum, but also continuously track the optimum over time. This section first describes the time-dependant optimization problem of identifying time-varying optimal contamination response protocols during an emergency. This includes a description of the time-dependant objective function of minimizing ultimate health impacts and optimization decision variables of contaminant flushing and public warning. It is followed by a brief introduction of four different groups of dynamic optimization algorithms. The particular dynamic evolutionary algorithm applied in this study is finally described in detail.

3.1. Time-varying Fitness

Emergency response decisions of hydrant operation and food-grade dye injection should be optimized for effective reduction of the public health consequences. The health impacts are expressed here mathematically as the ultimate total ingested mass (UTIM) of contaminant by all consumers during the course of a contamination incident:

$$\min \; UTIM = \sum_{i=1}^{N_p} \sum_{j=1}^{N_I} V_{i,j} \times C_{i,j} \tag{1}$$

where $N_p$ = size of population served by the WDS, $N_I$ = number of water ingestion events for each individual, $V_{i,j}$ = volume of water ingested by individual $i$ at ingestion event $j$, and $C_{i,j}$ = concentration of contaminant in water volume ingested by individual $i$ at ingestion event $j$. $V_{i,j}$ is assumed to be zero after an individual is alerted or gets sick. The clock time at which a consumer is alerted is determined by the EPANET toolkit extended-period hydraulic and quality run for simulation of dye propagation in WDS. Spatial and temporal distribution of contaminant concentration ($C_{i,j}$) is the output of EPANET toolkit extended-period run that simulates contaminant propagation in WDS. Therefore, every simulation for the determination of UTIM requires two EPANET toolkit runs; one for dye propagation and one for contaminant transport.

As described in the previous section, there exist multiple time-varying system parameters whose changing patterns are not known a priori because they are influenced by previous system states. Thus, Equation (1), which indicates the "predicted" value of ultimate health impacts at every stage of the emergency, represents a time-varying fitness function. Static optimization algorithms are insufficient for dealing with such changing objective functions. They need to be modified to adapt rapidly to changes in environment for generation of effective response protocols at every phase of the emergency. Obviously, the simplest approach to respond to a change in the environment is to consider each change as the emergence of a new optimization problem that needs be solved from scratch. Given sufficient time, this is obviously a feasible approach. However, the time available for re-optimization is normally short during an emergency and changes may occur frequently or even continuously. Moreover, this approach presumes that a change in the environment can be identified, which is not always true. Dynamic optimization techniques systematically reuse information from previously explored environments to accelerate optimization process in emerging environments.

Decision variables that should be optimized to minimize UTIM are the elements of response mechanisms of contaminant flushing and dye injection. In this paper, they include the locations for opening hydrants and injecting dye across the city and this will be called a response protocol. Hydrants of Mesopolis network are located on the intermediate nodes (i.e., zero-demand nodes that may connect to a series of terminal demand nodes) and are treated here as a possible location for contaminant discharge or food-grade dye injection. Total number of flushing and dye injection locations is set a priori, and depends on availability of labor and equipments.

3.2. Dynamic Optimization Algorithm

Evolutionary algorithms (EAs) resemble natural biological evolution, and since evolution is a continuous adaptation process in nature, they are promising candidates for tackling dynamic optimization problems [49]. To solve dynamic optimization problems, static EAs should be modified to adapt and recover from the changes during the evolution process. Major modifications in the static EAs are necessary for a timely adaptation to the changing environment to balance between convergence and exploration. Compared to static EAs, higher emphasis should be placed on exploration after a change occurs, so that the algorithm can react rapidly to the change and track the moving optimum. Different methods have been proposed to deal with this issue, which can be classified into four groups [22, 49]:

1) *Boost diversity after a change*: the EA is initially run in standard fashion. As soon as a change in the environment is identified, explicit strategies are implemented to generate diversity in the population. A common technique is hypermutation [50], where the mutation rate is significantly increased for a limited number of generations after the change event is detected and then decreased over time. A very high mutation rate essentially results in a re-initialization of the population, whereas a very low mutation rate does not boost sufficient diversity of the population.

2) *Maintain diversity throughout the run*: convergence is limited through constant diversification hoping that a diverse population is more promising to adapt to time-varying changes. Thermodynamic genetic algorithm [51], where the original objective function is replaced with an entropy-based value, and multiobjective-based method [52], where an artificial objective is used to promote diversity, are representatives of this approach. The multiobjective-based approach is used in this current study and is described in detail in Section 3.2 below.

3) *Memorize good solutions*: the algorithm retains good solutions from past generations. This strategy provides diversity and helps the algorithm retrieve the optimum in repetitive environments. Diploidy approach [53], where redundant representations are used to generate solutions, is a popular instance of memory-based approaches.

4) *Use multiple subpopulations*: the population is clustered into multiple subpopulations that evolve together to explore multiple promising regions of the decision making space. Some representative methods are self-organizing scouts [54] and finder-tracker multi-swarm particle swarm optimization [55].

This study employs the multiobjective-based diversity preservation approach, which has been demonstrated as a robust and efficient method by previous research [52, 56]. The main advantage of this technique is that it eliminates the need for defining a priori the proper diversity preservation parameter. The proper balance between convergence and exploration is systematically preserved during the process through treating diversity as a secondary (artificial) objective, which is optimized simultaneously with the main (true) optimization objective, as schematically illustrated in Figure 1.

The multiobjective-based diversity preservation approach used here *multiobjectivizes* [57] this classic single-objective optimization problem to a bi-objective optimization problem. Health impacts are minimized with simultaneous maximization of diversity hoping this increased diversity of the GA population helps tracking the time-dependant optimum. Once the *multiobjectivization* is performed, a traditional multiobjective optimization algorithm may be used to solve the constructed bi-objective optimization problem.

The artificial diversity-preservation metric may be mathematically expressed in different ways. Three following formulations are examined here:

1) *Distance from the nearest neighbor (DNN)*: The artificial objective for a solution $x_i$ is defined as the distance from $x_i$ to its nearest neighbor in the optimization decision space. Therefore, a pair of individuals that are very similar in the decision space will have a relatively poor artificial objective value, and thus diversity in population is encouraged over the search space.

$$DNN(x_i) = \min \ d(x_i, x_j), j = 1,...,N_{gp} \quad j \neq i \tag{2}$$

where $N_{gp}$ is the GA population size and $d(x_i, x_j)$ is the distance between two GA solutions (response protocols) in the optimization decision space. As each response protocols is defined by a string of locations for opening hydrants and injecting dye, sum of differences between geographical coordinates of these corresponding locations is used to calculate the distance between each response protocols pair. Selection of DNN as the artificial objective function implies that the model increases diversity through maximizing the minimum distance between solutions.

2) *Average distance from all solutions (ADS)*: Diversity is quantified as the average distance of $x_i$ to all other individuals in the GA population. This formulation prefers solutions at the edge of population to boost the spread of the population over the search space.

$$ADS(x_i) = \frac{1}{N_{gp}} \sum_{j=1}^{N_{gp}} d(x_i, x_j) \tag{4}$$

3) *Distance from the best solution in the current population (DBS)*: The diversity metric is expressed as the distance from $x_i$ to the current best solution in the GA population (with respect to the true objective function) to avoid any likely trap caused by local optima.

$$DBS(x_i) = d(x_i, x_{best}) \tag{3}$$

A major difference between DBS and the two other metrics is that it accounts for the dissimilarities in the objective space through comparing $x_i$ with the solution with minimum true objective space.

The bi-objective optimization problem defined by Equation (1) and any of Equations (2)-(4) may be solved using a multiobjective optimization algorithm. Non-dominated Sorting Genetic Algorithm II (NSGA-II) [58] is among the most popular algorithms for solving both classic water resources problems and emergent dynamic optimization problems [59-64]. NSGA-II is an elitist multiobjective evolutionary algorithm, which uses a fast strategy for non-dominated sorting and does not need any user-defined parameter for preservation of diversity in solutions. On the basis of the general concept of dominance in multiobjective optimization [65], response protocol $x_1$ dominates $x_2$ if two conditions are met: (1) $x_1$'s UTIM is smaller than or equal to $x_2$'s UTIM and $x_1$'s diversity is greater than or equal to $x_2$'s diversity, and (2) $x_1$'s UTIM is smaller than $x_2$'s UTIM or $x_1$'s diversity is greater than $x_2$'s diversity or both. The best response protocol in each time step that is reported to the emergency managers is the minimum-UTIM solution in that time step. The particular algorithm applied in this study uses simulated binary crossover [66], and polynomial mutation [65] for reproduction of offspring response protocols (i.e., contaminant flushing and dye injection locations). To explore new locations, reproduction operation is performed on coordinates of the parent solutions and the closest intermediate node in the WDS is chosen for the offspring solution.

## 4. Application Example

A virtual city, Mesopolis, with a population of nearly 147,000 is used to demonstrate the proposed dynamic framework [67]. In what follows, a general description of Mesopolis, considered contamination scenario and updates, and model settings are provided at first. Dynamic simulation-optimization model is then applied to identify and track time-dependant optimal response protocols. The performance of the three different diversity preservations metrics and response mechanisms of contaminant discharge and food-grade dye injection are evaluated and compared. A thorough analysis is then performed to understand how the different factors that contribute to the unpredictable time-varying system behavior affect the time-dependant optimal response protocols during an emergency individually and in combination.

4.1. WDS, Contamination Scenario, and Model Settings

Mesopolis is developed using a historical development timeline to incorporate as many attributes of a true city as possible. The WDS includes one reservoir, two water treatment plants (WTP), 13 tanks, 65 pumps, 876 hydrants, and 2,062 water mains and is shown in Figure 2. The East WTP distributes water across almost the entire city, while the West WTP supplies water only to western Mesopolis. Different types of consumers exist in Mesopolis such as residential, commercial, and industrial consumers. Water demands associated with different consumers fluctuate during the day. This temporal variability in demand is incorporated through defining distinct diurnal demand patterns for different consumers, as shown in Figure 3.

An arbitrary contamination emergency timeline is defined and used here as shown in Figure 4. It is assumed that the true contamination scenario is accurately identified at time 10:00, and perceived scenario is updated twice until then as described in Table 1. This updating practice is performed to analyze the effect of changes in perceived source attributes. Consideration of only two updates of perceived scenario attributes in this application example is for the sake of a more clear and understandable assessment and discussion. Without a need for any modifications, this model could be equally applicable when scenario is updated more frequently or even continuously.

The EPANET hydraulic time step is set to one hour. The simulation duration is 24 hours and the dynamic optimization model run starts after a 6-hour response delay after the injection starts, i.e. at 06:00. The underlying assumption here is that public exposure becomes comparatively negligible after the first day as the emergency managers take a comprehensive set of response actions including contaminant containment, discharge, public warning, etc.

The UTIM values associated with the time-varying perceived contamination scenario are reported in Table 2 for the conditions that consumers' reactions are ignored or considered. The observation that the ingested contaminant mass over injected load ratio is very small is explained by the fact that only an extremely small fraction of total urban water consumption is used for drinking. Table 2 indicates that smaller TD reduces the UTIM more because the individuals become sick earlier and cease drinking contaminated water, and this consequently decreases the ingested amount of contaminant by every individual and the entire population as a whole.

The values of UTIM listed in Table 2 indicate the ultimate value of total ingested mass (TIM) of contaminant by the entire population. In other words, as the optimization fitness function represented by Equation (1) also indicates, UTIM is equal to TIM at the end of emergency period, i.e. at 24:00. This projected accumulation of TIM over time for the contamination scenarios perceived at different time periods during the emergency is shown in Figure 5. Figures 5 (a), (b), and (c) show the projected time series of TIM at time periods of 06:00-08:00, 08:00-10:00, and 10:00-24:00, respectively, using the perceived scenarios associated with each period. It is observed that no contaminant mass is ingested until after hour 06:00 although the injection starts at 00:00. This is because the first daily water ingestion event happens at 07:00, as described in Section 2.3 above. The decreases in perceived injected contaminant mass from 300 kg to 250 kg during the emergency causes the UTIM and the overall slope of the TIM time series to decrease as observed in Figure 5 (a) to Figure 5 (c). The uncertainty bound of TIM due to the uncertainty in TD also widens as the perceived injected mass decreases.

The number of hydrants and dye injectors depend upon availability of resources and personnel during an emergency and are each set to 3 in this study. Sensitivity analyses on the effect of these two parameters on the effectiveness of response measures are provided in [13, 14]. Hydrants remain open for 5 hours once they are opened for flushing and are closed afterwards. The emitters section of the EPANET toolkit is utilized to realistically model the hydrants within the WDS. During the contaminant discharge period, the emitter discharge coefficient is set to 166.5 gpm/psi$^{0.5}$, which is associated with a 3-inch diameter connection fire hydrant. The amount of dye injected per each injector is set to 100 kg and the injection duration is 1 hour based upon conclusions made by [14] that shorter injection durations are more effective for reducing the impacts. The NSGA-II population size is set to 50. Crossover and mutation rates are 0.85 and 0.04, and simulated binary crossover distribution and polynomial mutation indices are set to 15 and 10, respectively.

All model runs in this application example are performed using a single core of a Dell Precision T5500 Workstation with Xeon® CPU running at 2.40 GHz and 6 GB RAM. Application of a supercomputer equipped with multi-level parallel processing software might significantly improve the computational efficiency, which helps identifying better response protocols at every time during an emergency. In this study, however, use of an ordinary desktop computer is preferred to account for the fact that most water utilities do not posses such advanced computational resources.

4.2. Optimization Results and Discussions

As described earlier, changes in emergency environment are due to multiple contributing factors including perceived scenarios updates, actions implemented by the utility operators, and reactions made by the alerted or sick consumers. The effect of these factors is neglected at first to investigate how performance of the identified optimal response protocol degrades as time passes. The effect of other contributing factors will be taken into account in later sections.

Fig. 6 indicates the UTIM time series associated with the best response protocol that is identified and updated recurrently by the dynamic simulation-optimization model as the emergency proceeds. This can be interpreted as the evolution plot of the optimal (i.e., minimum-UTIM) response protocol during the optimization process as the emergency proceeds. As the formulation of the true objective function (Equation (1)) explains, these time series indicate the ultimate value of total ingested mass after the emergency ends and not the value of total ingested mass during the emergency. This explains why the no-response UTIM time series shown in Figure 6 is a flat horizontal line. The response time series indicates how much UTIM can be reduced at every time during the event if the optimal response protocol identified at that time is implemented. Because the last daily water ingestion event happens at 18:00 and UTIM thus remains constant afterwards, the horizontal axis is cut at 18:00. The distinctive behavior of this evolution plot when compared with the common plots for static minimization problems (where the objective function exhibits a constant decreasing trend) is due to the fact that the global optimal response protocol here is a time-varying and generally degrading solution.

The horizontal axis in Figure 6, which denotes time, starts from 06:00 because the response delay is set here to 6 hours. In other words, the model run is started with a lapse of 6 hours after the contaminant intrusion starts to account for the time that it takes the emergency managers to learn about the contamination threat. Some contaminant mass has been thus ingested by the consumers before the managers respond. After the dynamic optimization model is run, the quality of optimal response protocols suggested by the model changes over time. Longer time allows the model to better explore the search domain and converge to the current global optimal solution more. This extended time, however, may reduce the effectiveness of the current global solution because of the prolonged exposure of the public to the contaminant and wider spread of the contamination. As time passes, public are more exposed to the contaminant and this degrades effectiveness of the current global solution. The observation that UTIM decreases during the first hour implies that model convergence outstrips global optimal solution degradation during the very early stages.

The UTIM time series follow a stepwise pattern with 1-hour increment jumps. This is because the hydraulic analysis time step is set to one hour in the EPANET hydraulic simulation model. A shorter time step favorably increases simulation accuracy and leads to a smoother curve but adversely increases computational time. Convergence of model to better solutions during the emergency may be clearly observed during every 1-hour time increment where UTIM gradually and consistently decreases. Hourly sudden jumps in UTIM indicate the reduction in response protocol performance as contaminant spreads more and the smooth decreasing trends are evidence of increased convergence to current optimum.

To analyze the performance of different diversity measures, separate model runs are performed using DNN, DBS, and ADS measures one for each complete run. The UTIM time series shown in Figure 6 visually demonstrate that DNN can most effectively reduce the UTIM at each time step. Numeric comparison may be also performed through calculating the area confined by each time series and the no-response line – a larger area indicates a better performance. The calculated areas associated with ADS, DBS, and DNN are 153.2, 129.4, and 189.5 gram-hours, respectively. Conclusively, both quantitative and visual comparisons indicate that DNN outperforms ADS and DBS and will be thus used for the remaining analyses in this study. The number of function evaluations is slightly different for different cases. Total number of function evaluations for the DNN, DBS, ADS cases in Figure 6 are equal to 14,950 (299 generations × 50 solutions), 14,850, and 15,000, respectively. Total runtime of the optimization process, however, is dictated by the emergency response time period and is thus the same (12 hours here) for all the cases in this paper regardless of their complexity. The average function evaluation time is thus approximately 2.9 seconds for DNN case, as an instance.

The performance of the hydrant operation and dye injection response mechanisms can be compared individually in the dynamic model. Figure 7 shows the time series for the minimum-UTIM response protocol offered by the optimization model using hydrant operation, dye injection, or both. The minimized health impacts reduce during the first hour and generally increase afterwards for all three cases. Visual inspection shows dye injection outperforms hydrant operation, and best performance is achieved when both strategies are implemented in combination. The areas confined by no-response and the response UTIM time series are 71.0, 147.8, and 189.5 gram-hours for time series corresponding to hydrant operation, dye injection, and both strategies, respectively. This quantitatively confirms the visual judgment outcome. This finding is specific to the particular response protocol settings described earlier (e.g., injected mass of dye and flushing duration) and may not be necessarily generalizable.

Besides the response execution delay, the effectiveness of optimal response protocols identified at every time step depends on the perceived scenario updates, previous response protocols executed by emergency managers, and changes in alerted or sick consumers' water demands. Fig. 8 shows how the UTIM time series changes as the effects of these contributing factors are taken into account individually or in combination.

Figure 8(a) indicates the UTIM time series when only the effects of perceived scenario updates, which are explained in Tables 1 and 2, are considered. Scenario updates significantly change the optimization search domain and the model adapts to these changes to track the moving optimum. This is evidenced by the observation that the drops and jumps in the no-response and response UTIM time series match each other. Further insight may be obtained through the illustration of moving optimum in the decision space as the perceived scenario is updated. Fig. 9 shows the spatial distribution of ultimate ingested contaminant mass at 07:59 (i.e. right before the first update) and 11:05 (i.e. two hours after the first update). At 07:59 (when UTIM is 195.4 g), the East WTP is perceived as the contaminant injection location and this explains why the projected impact area is larger than 11:05 (when UTIM is 186.7 g) when the perceived injection location is Node IN654, which is an intermediate node. Figure 9 also shows that the currently identified optimal response protocol significantly changes in response to the perceived scenario updates. Suggested optimal locations for contaminant discharge and dye injection noticeably match the projected impact areas at each of considered clock times.

As it was shown in Figure 4, the managers execute the current best minimum-UTIM response protocols at clock times 09:00 and 12:00 for the emergency timeline example application presumed here. Consequences of these actions are manifested in the UTIM time series shown in Figure 8(b) by a decreasing trend in UTIM starting from each execution time. Observation of extended decrease rather than a sudden drop is because the response protocols are implemented over a period of time – dye is injected over 1 hour and hydrants are open for 5 hours. The adaptive model takes into account the effect of these actions when later actions are evaluated. This means the response protocol is now optimized given one or more optimal response protocols are already executed or are being implemented. In other words, the model superposes current response protocols onto the previously executed ones. The reason no-response curve is different from Fig. 8(a) is that the effects of perceived scenario updates are not considered in Fig. 8(b).

Compared to Figures 6 and 7, Figure 8(b) shows that the originally increasing trend of the time series transforms into a flat line after the response protocols are executed at 09:00 and 12:00. This is because the contaminant concentrations are much reduced in the WDS due to previous contaminant discharge and public warning actions that together reduce public exposure risk. This also may imply that the increased quality of solutions due to convergence now neutralizes the degradation in solutions performance due to the prolonged contaminant spread time.

Figure 8(c) indicates the UTIM time series when the effect of changes in consumers' water consumption behavior, i.e. ceasing drinking tap water and reducing water demands, is taken into account. UTIM time series are presented for three different TD estimations for both when response is taken or not. In general, the UTIM time series are very similar in trend to when the demand changes effects are not considered (Figure 7). Consideration of changes in consumers' behavior lowers the ultimate health impacts because ceasing to drink tap water reduces the possibility of ingesting more amount of the contaminant. This is also in agreement with the observation that setting TD to smaller values leads to a larger reduction in UTIM. When TD is smaller, the ingested contaminant mass exceeds TD for a larger number of consumers faster. This subsequently reduces UTIM as more consumers stop drinking contaminated water. Reduction in water use by different residential, commercial, and industrial consumers also alters WDS hydraulics, and thus contaminant and dye propagation. This effect consequently changes the global optimal response protocol at different times during the event.

The UTIM time series for the most realistic representation of the system dynamics is shown in Figure 8(d). This time series accounts for all the contributing factors including temporal variability of perceived contaminant spread, response actions taken by the emergency managers, and water ingestion and demand changes by the alerted or sick consumers. This inclusive time series is understandably similar in trend and intensity to a time series that one may envision through the superposition of the time series associated with each of the contributing factors individually (Figures 8(a)-(c)). Sudden changes in UTIM at times 08:00 and 10:00 are due to updates made to the perceived contamination scenario at these times. Implementation of optimal response protocols at 09:00 and 12:00 reduces the UTIM as expected. This is accompanied by general decrease in UTIM throughout the entire time period as more alerted and sick consumers stop drinking contaminated water.

## 5. Conclusions and Future Work

An innovative modeling approach to contamination threat management in drinking water networks was presented by taking into account the unpredictable time-varying factors such as perceived contaminant source attributes and the feed-back mechanisms between the hydraulic network, consumers, and managers. Through the incorporation of these random factors, the dynamic simulation provides a more realistic picture of the uncertain contamination emergency environment than the static approaches that presume complete knowledge about present and future conditions. Once integrated with a dynamic optimization algorithm that methodologically preserves adaptation, the resulting dynamic simulation-optimization model was able to identify and track time-varying optimal health-protection measures to serve the needs of utility operators during the course of an emergency. Demonstration of the proposed adaptive scheme on a virtual city using a regular desktop computer indicates this model can be used as a helpful decision support tool by water utilities to reduce water contamination health impacts.

There is no claim that the response protocols identified by the proposed dynamic optimization scheme are globally optimal as with any static or dynamic EA-based model. Nevertheless, because EAs mimic natural evolution and evolution is an adaptation process in nature, they are fitting candidates for solving nonlinear optimization problems in changing environments. Future studies may examine performance of other types of dynamic EA-based schemes described in this paper for a more efficient optimization of response protocols during a contamination emergency.

Hypothetical contamination events were used in this study to investigate the applicability of proposed adaptive framework for protecting public health against contamination threats. Although complete validation of the model is not feasible because data is not available about real intentional WDS contamination incidents, the proposed modeling framework is useful for analyzing what-if scenarios to enhance the state of preparedness. Knowledge about the degradation of the optimal response protocols performance over time for different intrusion scenarios can be obtained and illustrated as time series of optimal reduction in ultimate health impacts. These informative plots are useful for estimation of the added value of rapid response once an intrusion scenario occurs and identification of critical scenarios. They can be utilized to guiding risk mitigation and emergency preparedness decisions.

The multiobjective-based dynamic optimization algorithm used in this study defines and maximizes an artificial objective function for preserving diversity to enhance adaptability to changing environments. The performance of three different diversity preservation measures (distance from the nearest neighbor, average distance from all solutions, and distance from the best solution in the current population) were investigated. Visual and quantitative comparisons indicated the distance from the nearest neighbor measure would provide higher quality response protocols during an emergency.

Relatively simple rules were used in this study to simulate the complex interactions between the hydraulic system and consumers that change the emergency environment conditions. Advanced agent-based models and sociotechnical simulation frameworks [15, 43] may be used to better capture system dynamics, such as word-of-mouth communication between consumers and their mobility, and provide a more realistic picture of an emergency environment. Due to the increased computational intensity, however, the applicability of such compute-intensive complex models during the course of an emergency is subject to the availability of more advanced hardware and software computational resources than the regular desktop computers used in this study.

The adaptive simulation-optimization model described in this paper is intended to be a major component of a cyberinfrastructure for drinking water network contamination threat management. This cyberinfrastructure will include multiple modules such as real-time SCADA data processing schemes, adaptive contaminant source identification models, graphical user interfaces, among others. Work is underway to construct and embed these modules. This adaptive cyberinfrastructure can be used both as a realistic emergency simulator for training the water utility operators during the preparedness phase and also as a valuable decision support system for an intelligent and informed emergency response in the event of a real contamination threat.

## 6. Acknowledgements

This material is based upon work supported by the U.S. National Science Foundation under Grant Number CMMI-0927739. Any opinions, findings, or recommendations expressed in this paper are those of the authors and do not necessarily reflect the views of the sponsors.

## 7. References

[1] Staudinger, T. J., England, E. C., & Bleckmann, C. (2006). Comparative analysis of water vulnerability assessment methodologies. *Journal of Infrastructure Systems*, 12(2), 96–106.

[2] Reynolds, K. A., Mena, K. D., & Gerba, C. P. (2008). Risk of waterborne illness via drinking water in the United States. *Reviews of Environmental Contamination and Toxicology*, 192, 117–158.

[3] United States Congress (2002). Public health security and bioterrorism preparedness and response act of 2002. Public Law No. 107-188, H.R. 3448, 107th Congress, Government Printing Office, Washington, D.C.

[4] Lindell, M. K., Prater, C. S., & Perry, R. W. (2006). Fundamentals of emergency management. Federal Emergency Management Agency, Hyattsville, MD.

[5] Baranowski T. M., & LeBoeuf, E. J. (2008). Consequence management utilizing optimization. *Journal of Water Resources Planning and Management*, 134(4), 386–394.

[6] Poulin, A., Mailhot, A., Grondin, P., Delorme, A., Periche, A., & Villeneuve, J. P. (2008). Heuristic approach for operational response to drinking water contamination. *Journal of Water Resources Planning and Management*, 134(5), 457–265.

[7] Shafiee, M. E., & Berglund, E. Z. (2014) Real-time guidance for hydrant flushing using sensor-hydrant decision tree. *Journal of Water Resources Management and Planning*, in press.

[8] Shafiee, M., & Zechman, E. M. (2011). Sociotechnical simulation and evolutionary algorithm optimization for routing siren vehicles in a water distribution contamination event. *Genetic and Evolutionary Computation Conference*, Dublin, Ireland.

[9] Shafiee, M. & Zechman, E. (2012). Sociotechnical Simulation for Evaluating Adaptive Threat Response Actions for Water Distribution Contamination Events. *World Environmental and Water Resources Congress 2012*: pp. 3285–3291.

[10]  Shafiee, M. (2013). Integrating an agent-based model with a multi-objective evolutionary algorithm to warn consumers in a water contamination event using emergency vehicles. *World Environmental and Water Resources Congress 2013*: pp. 2550–2558.

[11]  Preis, A., & Ostfeld, A. (2008). Multiobjective contaminant response modelling for water distributions systems security. *Journal of Hydroinformatics*, 10(4), 267–274.

[12]  Alfonso, L., Jonoski, A., & Solomatine, D. (2009). Multiobjective optimization of operational responses for contaminant flushing in water distribution networks. *Journal of Water Resources Planning and Management*, 136(1), 48–58.

[13]  Rasekh, A., & Brumbelow, K. (2014). Drinking water distribution systems contamination management to reduce public health impacts and system service interruptions. *Environmental Modelling & Software*, 51, 12–25.

[14]  Rasekh, A., Brumbelow, K., & Lindell, M. (2014). Water as warning medium: food-grade dye injection for drinking water contamination emergency response. *Journal of Water Resources Planning and Management*, 140(1), 12–21.

[15]  Zechman, E. M. (2011). Agent-based modeling to simulate contamination events and evaluate threat management strategies in water distribution systems. *Risk Analysis*, 31(5), 758–772.

[16]  Shafiee, M. E., & Zechman, E. M. (2013). An agent-based modeling framework for sociotechnical simulation of water distribution contamination events. *Journal of Hydroinformatics*, 15(3), 862–880.

[17]  Branke, J. (2001). Evolutionary optimization in dynamic environments. kluwer Academic Publishers, Norwell, MA.

[18]  Del Amo, I. G., Pelta, D. A., González, J. R., & Masegosa, A. D. (2012). An algorithm comparison for dynamic optimization problems. *Applied Soft Computing*, 12(10), 3176–3192.

[19]  Besbes, O., & Zeevi, A. (2009). Dynamic pricing without knowing the demand function: Risk bounds and near-optimal algorithms. *Operations Research*, 57(6), 1407–1420.

[20]  Liu, L., Ranjithan, R., & Mahinthakumar, G. (2011). Contamination source identification in water distribution systems using an adaptive dynamic optimization procedure. *Journal of Water Resources Planning and Management*, 137(2), 183–192.

[21] Khouadjia, M. R., Sarasola, B., Alba, E., Jourdan, L., & Talbi, E. G. (2012). A comparative study between dynamic adapted PSO and VNS for the vehicle routing problem with dynamic requests. *Applied Soft Computing*, 12(4), 1426–1439.

[22] Bui, L. T., Michalewicz, Z., Parkinson, E., & Abello, E. M. (2012). Adaptation in dynamic environments: A case study in mission planning. *IEEE Transactions on Evolutionary Computation*, 16(2), 190–209.

[23] Kong, W., Chai, T., & Yang, S. (2013). A hybrid evolutionary multiobjective optimization strategy for the dynamic power supply problem in magnesia grain manufacturing. *Applied Soft Computing*, 13(5), 2960–2969.

[24] Connolly, J. F., Granger, E., & Sabourin, R. (2012). Dynamic multi-objective evolution of classifier ensembles for video face recognition. *Applied Soft Computing*, 13(6), 3149–3166.

[25] Mavrovouniotis, M., & Yang, S. (2013). Ant colony optimization with immigrants schemes for the dynamic travelling salesman problem with traffic factors. *Applied Soft Computing*, 13(10), 4023–4037.

[26] Rasekh, A., and Brumbelow, K. (2013). Probabilistic analysis and optimization to characterize critical water distribution system contamination scenarios. *Journal of Water Resources Planning and Management*, 139(2), 191–199.

[27] Murray, S., McBean, E., & Ghazali, M. (2012). Real-time water quality monitoring: assessment of multisensor data using Bayesian belief networks. *Journal of Water Resources Planning and Management*, 138(1), 63–70.

[28] Hart, D.B., McKenna, S.A., Klise, K.A., Wilson, M.P., & Murray, R. (2009). CANARY user's manual and software upgrades. U.S. Environmental Protection Agency, Washington, D.C., EPA/600/R-08/040A.

[29] Shahrokh Esfahani, M., & Dougherty, E. R. (2014). Incorporation of Biological Pathway Knowledge in the Construction of Priors for Optimal Bayesian Classification. *IEEE/ACM Transactions on Computational Biology and Bioinformatics*, 11(1), 202–218.

[30]  Shahrokh Esfahani, M., & Dougherty, E. R. (2014). Effect of separate sampling on classification accuracy. *Bioinformatics*, 30(2), 242–250.

[31]  Mahinthakumar, K., von Laszewski, G., Ranjithan, R., Uber, J., Harrison, K., Sreepathi, S., & Zechman, E. (2006). An adaptive cyberinfrastructure for threat management in urban water distribution systems. In Alexander et al. (Eds). Lecture Notes in Computer Science, vol 3993 (pp. 401–408). Springer-Verlag, Berlin.

[32]  Perry, R.W., & Lindell, M.K., (2007). Emergency planning. John Wiley, Hoboken NJ.

[33]  United States Environmental Protection Agency (USEPA) (2004a). Response protocol toolbox: Planning for and responding to drinking water contamination threats and incidents—Module 6: Remediation and Recovery Guide. USEPA, Washington, D.C.

[34]  Haxton, T., Murray, R. & Klise, K. (2012). Examining the application of modeling tools to identify effective flushing locations. *World Environmental and Water Resources Congress 2012*: pp. 3071–3081.

[35]  U.S. Food and Drug Administration (2013). Summary of Color Additives for Use in the United States in Foods, Drugs, Cosmetics, and Medical Devices. Available online: http://www.fda.gov/forindustry/coloradditives/coloradditiveinventories/ucm115641.htm

[36]  Davis, M. J., & Janke, R. (2009). Development of a probabilistic timing model for the ingestion of tap water. *Journal of Water Resources Planning and Management*, 135(5), 397–405.

[37]  United States Environmental Protection Agency (USEPA) (2004b). Estimated per capita water ingestion in the United States. EPA-822-R-00-001, USEPA, Washington, D.C.

[38]  White, J. W. (1999). Hazards of short-term exposure to arsenic-contaminated soil. Office of Environmental Health Assessment Services, Olympia, WA.

[39]  Lopez-de-Alba, P. L., Lopez-Martinez, L., Cerda, V., & De-Leon-Rodriguez, L. M. (2001). Simultaneous determination of tartrazine, sunset yellow and allura red in commercial soft drinks by multivariate spectral analysis. *Quimica Analitica*, 20(2), 63–72.

[40]  Vickers, A. (2001). Handbook of water use and conservation. WaterPlow Press, Amherst, MA.

[41] Zechman, E. M., Brumbelow, K., Lindell, M., Mumpower, J., Rasekh, A. & Shafiee, M (2011). Agentbased modeling for planning emergency response to contamination emergencies in water utilities. *NSF CMMI Engineering Research and Innovation Conference*, Atlanta, GA.

[42] Rasekh, A., Brumbelow, K., & Zechman, E. M. (2010). WDS vulnerability analysis: focusing on random factors, consumer behavior, and system dynamics in contamination events. *World Environmental and Water Resources Congress 2010*: pp. 4357–4363.

[43] Rasekh, A., Shafiee, M. E., Zechman, E., & Brumbelow, K. (2014). Sociotechnical risk assessment for water distribution system contamination threats. *Journal of Hydroinformatics*, 16(3), 531–549.

[44] Rossman L. A. (2000). EPANET 2.0, user's manual. National Risk Management Research Laboratory, U.S. EPA, Cincinnati, OH.

[45] Wu, Z. Y., Wang, R. H., Walski, T. M., Yang, S. Y., Bowdler, D., & Baggett, C. C. (2006). Efficient pressure dependent demand model for large water distribution system analysis. *Water Distribution System Analysis. Water Distribution Systems Analysis Symposium 2006*: pp. 1–15.

[46] Giustolisi, O., & Laucelli, D. (2010). Water distribution network pressure-driven analysis using the enhanced global gradient algorithm (EGGA). *Journal of Water Resources Planning and Management*, 137(6), 498–510.

[47] Shang, F., Uber, J., & Rossman, L. A. (2008). EPANET multi-species extension user's manual. U.S. Environmental Protection Agency, EPA/600/C-10/002, Washington, DC.

[48] Machell, J., Mounce, S. R., & Boxall, J. B. (2010). Online modelling of water distribution systems: a UK case study. Drinking Water Engineering and Science, 3(1), 21-27.

[49] Jin, Y., & Branke, J. (2005). Evolutionary optimization in uncertain environments: A survey. *IEEE Transactions on Evolutionary Computation*, 9(3), 303–317.

[50] Cobb, H. G. (1990). An investigation into the use of hypermutation as an adaptive operator in genetic algorithms having continuous, time-dependent nonstationary environments. Naval Research Libratory, Washington, DC.

[51] Mori, N., Imanishi, S., Kita, H., & Nishikawa, Y. (1997). Adaptation to changing environments by means of the memory-based thermodynamical genetic algorithm. *Proceedings of the 7th International Conference on Genetic Algorithms*, East Lansing, MI.

[52] Bui, L. T., Nguyen, M., Branke, J., & Abbass, H. A. (2008). Tackling dynamic problems with multiobjective evolutionary algorithms. In Knowles, J., Corne, D., & Deb, K. (Eds.) Multiobjective problem solving from nature (pp. 77–92). Springer, Berlin.

[53] Goldberg, D. E., & Smith, R. E. (1987). Nonstationary function optimization using genetic algorithms with dominance and diploidy. *2nd International Conference on Genetic Algorithms*, Cambridge, MA.

[54] Branke, J., Kaußler, T., Schmidt, C., & Schmeck, H. (2000). A multipopulation approach to dynamic optimization problems. *4th International Conference on Adaptive Computing in Design and Manufacture*, Devon, UK.

[55] Yazdani, D., Nasiri, B., Sepas-Moghaddam, A., & Meybodi, M. R. (2013). A novel multi-swarm algorithm for optimization in dynamic environments based on particle swarm optimization. *Applied Soft Computing*, 13(4), 2144–2158.

[56] Toffolo, A., & Benini, E. (2003). Genetic diversity as an objective in multi-Objective evolutionary algorithms. *Evolutionary Computation*, 11(2), 151–167.

[57] Handl, J., Lovell, S. C., & Knowles, J. (2008). Multiobjectivization by decomposition of scalar cost functions. *10th international conference on Parallel Problem Solving from Nature*, Dortmund, Germany.

[58] Deb, K., Pratap, A., Agarwal, S., & Meyarivan, T. (2002). A fast and elitist multi-objective genetic algorithm NSGA-II. *IEEE Transactions on Evolutionary Computations*, 6(2), 182–197.

[59] Nicklow, J., Reed, P., Savic, D., Dessalegne, T., Harrell, L., Chan-Hilton, A., Karamouz, M., Minsker, B., Ostfeld, A., Singh, A., & Zechman, E. (2010). State of the Art for Genetic Algorithms and Beyond in Water Resources Planning and Management. *Journal of Water Resources Planning and Management*, 136(4), 412–432.

[60] Bureerata, S., Sriworamas, K. (2013). Simultaneous topology and sizing optimization of a water distribution network using a hybrid multiobjective evolutionary algorithm. *Applied Soft Computing*, 13(8), 3693–3702.

[61] Afshar, A., & Takbiri, Z. (2012). Fusegates selection and operation: simulation–optimization approach. *Journal of Hydroinformatics*, 14(2), 464–477.

[62] Afshar, A., Rasekh, A., & Afshar, M. H. (2009). Risk-based optimization of large flood-diversion systems using genetic algorithms. *Engineering Optimization*, 41(3), 259–273.

[63] Rasekh, A., Afshar, A., & Afshar, M. H. (2010). Risk-cost optimization of hydraulic structures: methodology and case study. *Water Resources Management*, 24(11), 2833–2851.

[64] Takbiri, Z., & Afshar, A. (2012). Multi-objective optimization of fusegates system under hydrologic uncertainties. *Water resources management*, 26(8), 2323–2345.

[65] Deb, K. (2001). Multiobjective optimization using evolutionary algorithms. Wiley, Chichester.

[66] Deb, K., & Agrawal, R. B. (1995). Simulated binary crossover for continuous search space. *Complex Systems*, 9(2), 115–148.

[67] Johnston, G., Brumbelow, K., 2008. Developing Mesopolis: a 'virtual city' for research in water distribution systems and interdependent infrastructures. Available online: https://ceprofs.tamu.edu/kbrumbelow/Mesopolis/Developing%20Mesopolis.pdf

**Table 1.** Time-varying perceived contamination scenario attributes

| Period | Injection location | Load (kg) | Demand multiplier | Start time | Duration (hr) |
|---|---|---|---|---|---|
| 06:00 to 08:00 | East WTP | 300 | 1.00 | 00:00 | 5 |
| 08:00 to 10:00 | IN0654 | 280 | 1.00 | 00:00 | 4 |
| 10:00 to 24:00 | IN0654 | 250 | 1.00 | 00:00 | 5 |

**Table 2.** Ultimate health impacts associated with time-varying perceived contamination scenario. The results are for the entire simulation period of 24 hours.

| Period | UTIM (consumers do not react) (grams) | UTIM (consumers react) (grams) | | |
|---|---|---|---|---|
| | | Min TD | Average TD | Max TD |
| 06:00 to 08:00 | 220.0 | 211.8 | 217.6 | 220.0 |
| 08:00 to 10:00 | 186.2 | 156.6 | 170.5 | 178.7 |
| 10:00 to 24:00 | 192.7 | 157.5 | 167.6 | 191.7 |

**Figure captions**

**Figure 1** Diversification of GA solutions in multiobjective-based dynamic optimization algorithm for methodological balance between exploitation and exploration. Non-dominated response protocols are illustrated as filled circles.

**Figure 2** Water distribution system of Mesopolis virtual city

**Figure 3** Water consumption patterns for three different types of consumers in Mesopolis virtual city

**Figure 4** The example emergency timeline used for model application

**Figure 5** Accumulation of TIM over time projected based on the contamination scenario perceived at different time periods (see Tables 1 and 2). The ultimate projected value of TIM at the end of emergency period, i.e. 24:00, denotes UTIM (see Equation (1) and Table 2).

**Figure 6** The UTIM time series for different diversity measures ignoring the effects of perceived scenario update, response protocol execution, and consumers' demand changes

**Figure 7** The UTIM time series for different response strategies ignoring the effects of perceived scenario updates, response protocol executions, and consumers' demand changes

**Figure 8** The UTIM time series for combined hydrant and dye injection operation and considering the effects of (a) perceived scenario updates only (see Table 1), (b) response protocol executions only (see Figure 4), (c) consumers' reactions only, and (d) all factors together (using average TD estimate)

**Figure 9** Spatial distribution of projected ultimate ingested contaminant mass associated with the optimal response protocols at (a) 07:59 and (b) 11:05. Effect of perceived scenario updates is considered only (see Table 1 and Fig. 8(a)). Pipes are removed from the map for better readability

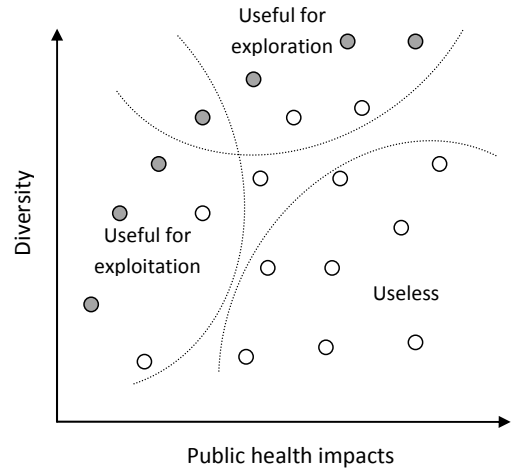

**Figure 1** Diversification of GA solutions in multiobjective-based dynamic optimization algorithm for methodological balance between exploitation and exploration. Non-dominated response protocols are illustrated as filled circles.

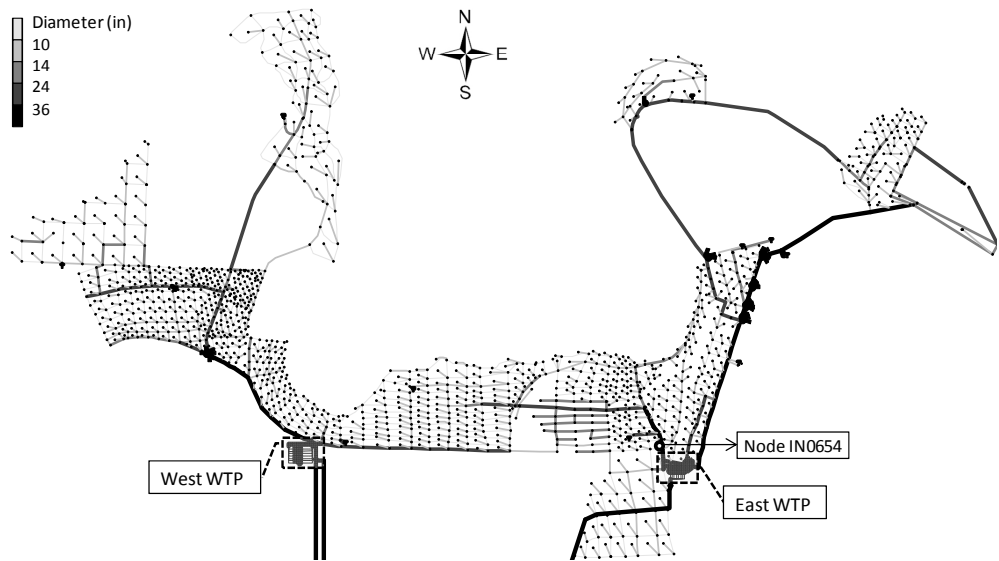

**Figure 2** Water distribution system of Mesopolis virtual city

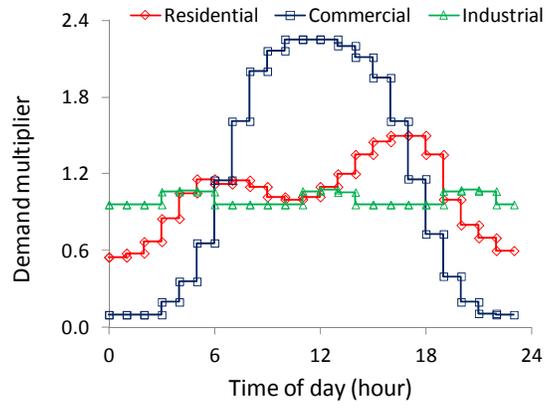

**Figure 3** Water consumption patterns for three different types of consumers in Mesopolis virtual city

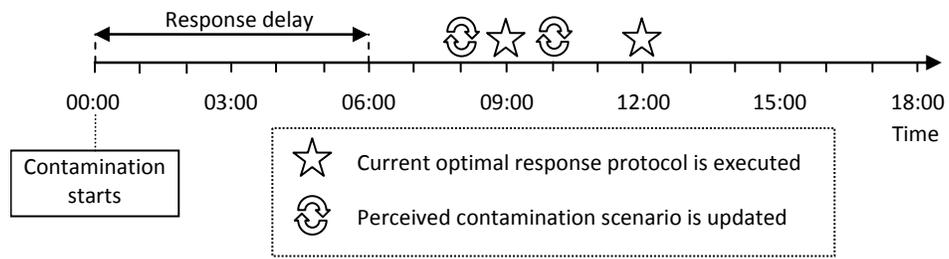

**Figure 4** The example emergency timeline used for model application

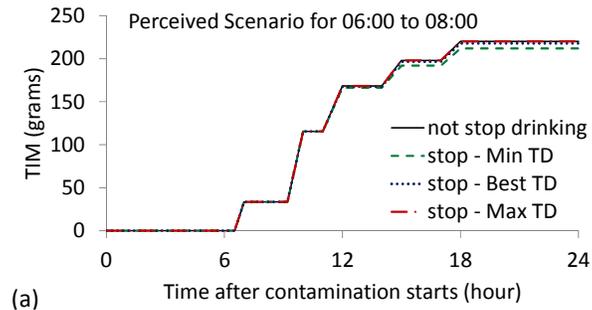

(a)

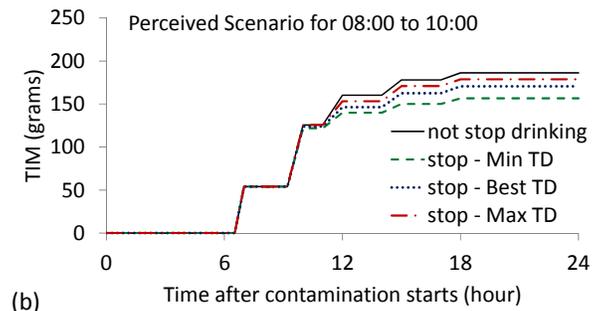

(b)

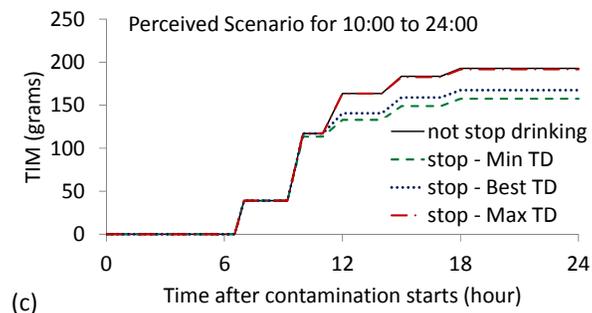

(c)

**Figure 5** Accumulation of TIM over time projected based on the contamination scenario perceived at different time periods (see Tables 1 and 2). The ultimate projected value of TIM at the end of emergency period, i.e. 24:00, denotes UTIM (see Equation (1) and Table 2).

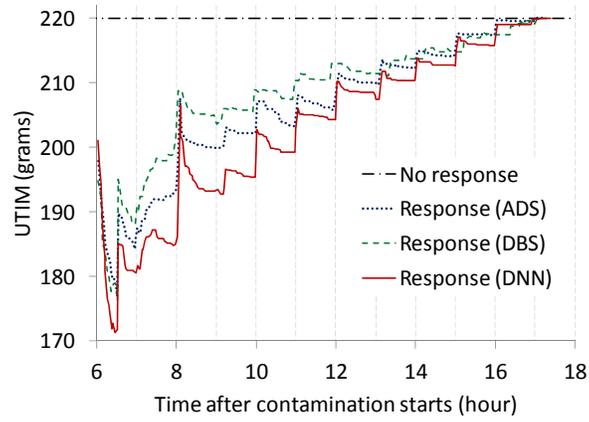

**Figure 6** The UTIM time series for different diversity measures ignoring the effects of perceived scenario update, response protocol execution, and consumers' demand changes

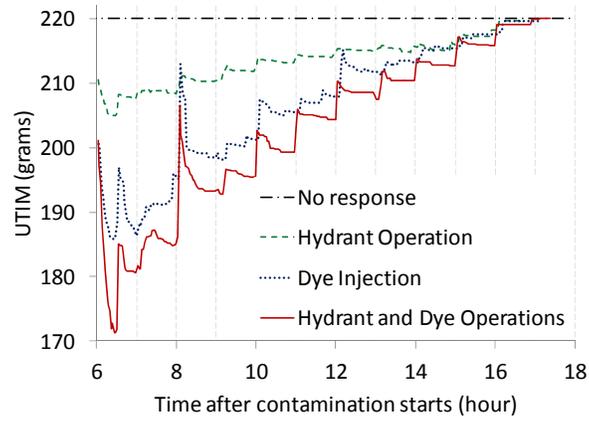

**Figure 7** The UTIM time series for different response strategies ignoring the effects of perceived scenario updates, response protocol executions, and consumers' demand changes

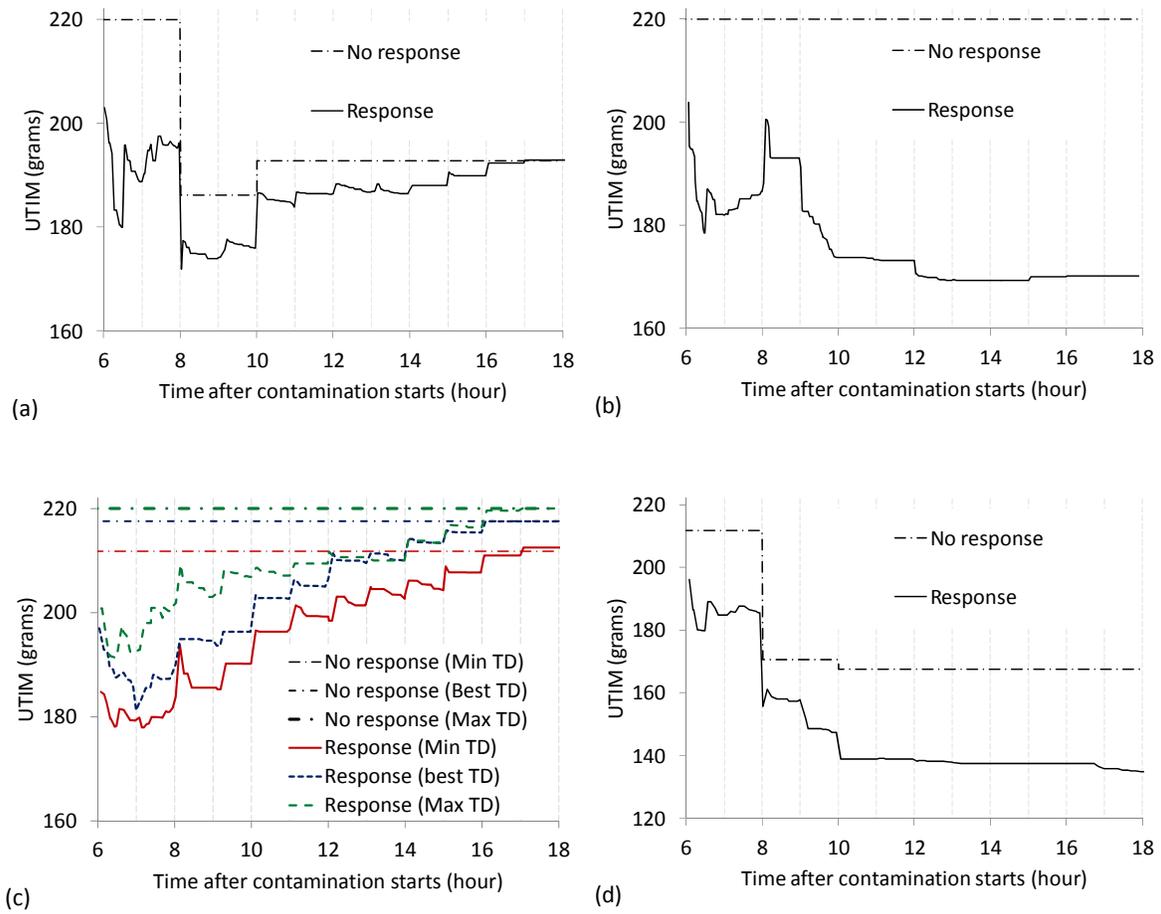

**Figure 8** The UTIM time series for combined hydrant and dye injection operation and considering the effects of (a) perceived scenario updates only (see Table 1), (b) response protocol executions only (see Figure 4), (c) consumers' reactions only, and (d) all factors together (using average TD estimate)

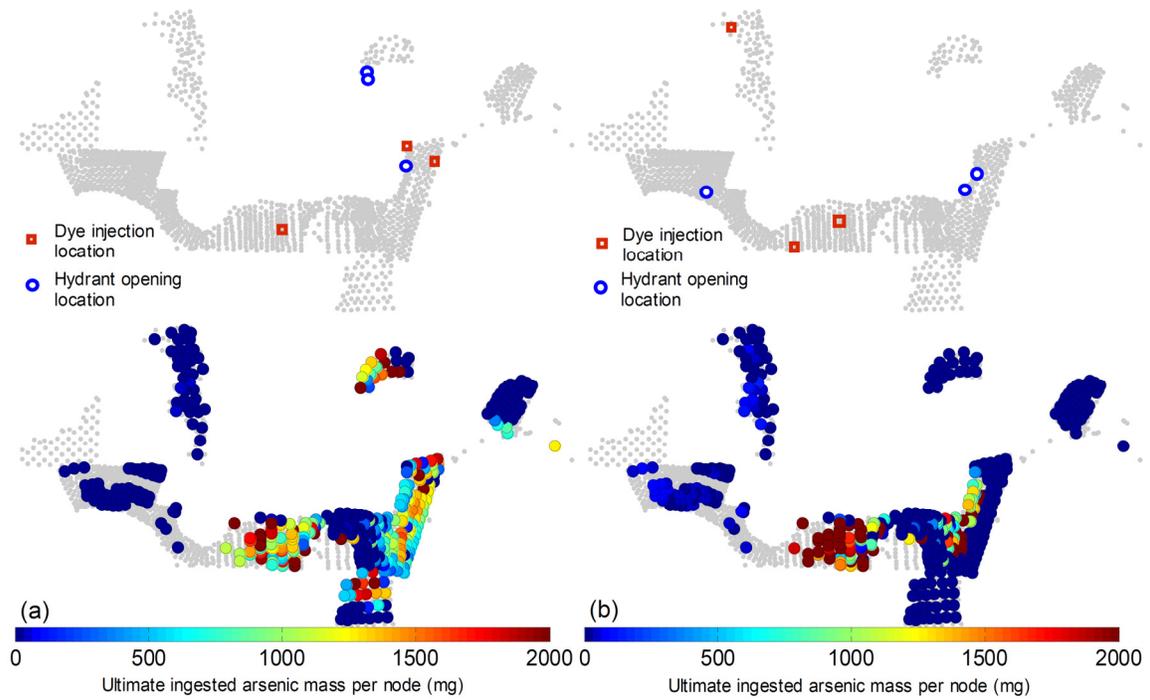

**Figure 9** Spatial distribution of projected ultimate ingested contaminant mass associated with the optimal response protocols at (a) 07:59 and (b) 11:05. Effect of perceived scenario updates is considered only (see Table 1 and Fig. 8(a)). Pipes are removed from the map for better readability.